\documentclass[%
 aip,
hidelinks,
 amsmath,amssymb,
 reprint,%
]{revtex4-1}

\usepackage{graphicx}
\usepackage{dcolumn}
\usepackage{bm}

\usepackage[utf8]{inputenc}
\usepackage[T1]{fontenc}
\usepackage{mathptmx}
\usepackage{etoolbox}
\usepackage{orcidlink}
\makeatletter
\def\@email#1#2{%
 \endgroup
 \patchcmd{\titleblock@produce}
  {\frontmatter@RRAPformat}
  {\frontmatter@RRAPformat{\produce@RRAP{*#1\href{mailto:#2}{#2}}}\frontmatter@RRAPformat}
  {}{}
}%
\makeatother
\begin{document}
\preprint{AIP/123-QED}

\title{Radio Frequency from Optical with Instabilities below $10^{-15}$- Generation\\and Measurement}
\author{A.~Hati\orcidlink{0000-0001-8363-8648}}
\affiliation{ 
\mbox{National Institute of Standards and Technology, 325 Broadway, Boulder, Colorado 80305, USA}
}
\email{archita.hati@nist.gov}

\author{M.~Pomponio\orcidlink{0000-0001-9158-0239}}
\affiliation{ 
\mbox{National Institute of Standards and Technology, 325 Broadway, Boulder, Colorado 80305, USA}
}
\affiliation{ 
\mbox{Electrical, Computer and Energy Engineering, University of Colorado, Boulder, Colorado 80309, USA}
}

\author{N.~V.~Nardelli\orcidlink{0000-0002-1558-5182}}
\affiliation{ 
\mbox{National Institute of Standards and Technology, 325 Broadway, Boulder, Colorado 80305, USA}
}

\author{T.~Grogan\orcidlink{0009-0003-3008-1876}}
\affiliation{ 
\mbox{National Institute of Standards and Technology, 325 Broadway, Boulder, Colorado 80305, USA}
}
\affiliation{ 
Department of Physics, University of Colorado, Boulder, Colorado 80309, USA
}

\author{K.~Kim\orcidlink{0000-0001-5428-6700}}
\affiliation{ 
JILA,University of Colorado, Boulder, Colorado 80309, USA
}
\affiliation{ 
\mbox{National Institute of Standards and Technology, 325 Broadway, Boulder, Colorado 80305, USA}
}
\affiliation{ 
Department of Physics, University of Colorado, Boulder, Colorado 80309, USA
}

\author{D.~Lee\orcidlink{0000-0002-5779-0010}}
\affiliation{ 
JILA,University of Colorado, Boulder, Colorado 80309, USA
}
\affiliation{ 
\mbox{National Institute of Standards and Technology, 325 Broadway, Boulder, Colorado 80305, USA}
}
\affiliation{ 
Department of Physics, University of Colorado, Boulder, Colorado 80309, USA
}
\author{J.~Ye\orcidlink{0000-0003-0076-2112}}
\affiliation{ 
JILA,University of Colorado, Boulder, Colorado 80309, USA
}
\affiliation{ 
\mbox{National Institute of Standards and Technology, 325 Broadway, Boulder, Colorado 80305, USA}
}
\affiliation{ 
Department of Physics, University of Colorado, Boulder, Colorado 80309, USA
}

\author{T.~M.~Fortier\orcidlink{0000-0003-2282-4986}}
\affiliation{ 
\mbox{National Institute of Standards and Technology, 325 Broadway, Boulder, Colorado 80305, USA}
}

\author{A.~Ludlow\orcidlink{0009-0001-8584-889X}}
\affiliation{ 
\mbox{National Institute of Standards and Technology, 325 Broadway, Boulder, Colorado 80305, USA}
}
\affiliation{ 
\mbox{Electrical, Computer and Energy Engineering, University of Colorado, Boulder, Colorado 80309, USA}
}
\affiliation{ 
Department of Physics, University of Colorado, Boulder, Colorado 80309, USA
}

\author{C.~W.~Nelson\orcidlink{0000-0001-5293-5113}}
\affiliation{ 
\mbox{National Institute of Standards and Technology, 325 Broadway, Boulder, Colorado 80305, USA}
}
\date{\today}

\begin{abstract}
This paper presents a frequency synthesis that achieves exceptional stability by transferring optical signals to the radio frequency (RF) domain at 100~MHz. We describe and characterize two synthesis chains composed of a cryogenic silicon cavity-stabilized laser at 1542~nm and an ultra-low expansion (ULE) glass cavity at 1157~nm, both converted to 10~GHz signals via Ti:Sapphire and Er/Yb:glass optical frequency combs (OFCs). The 10~GHz microwave outputs are further divided down to 100~MHz using a commercial microwave prescaler, which exhibits a residual frequency instability of $\sigma_y(1~\text{s})<10^{-15}$ and low $10^{-18}$ level at a few thousand seconds. Measurements are performed using a newly developed custom ultra-low-noise digital measurement system and are compared to the carrier-suppression technique. The new system enables high-sensitivity evaluation across the entire synthesis chain, from the optical and microwave heterodynes as well as the direct RF signals. Results show an absolute instability of $\sigma_y(1~\text{s})~\approx~4.7\times10^{-16}$ at 100~MHz. This represents the first demonstration of such low instability at 100~MHz, corresponding to a phase noise of -140~dBc/Hz at a 1~Hz offset and significantly surpassing earlier systems. These advancements open new opportunities for precision metrology and timing systems.
\end{abstract}
\maketitle
\section{\label{sec:level1}Introduction}
Generating extremely stable radio frequency (RF) signals from optical sources is an important capability that benefits high-precision radar, navigation, communication systems, and metrology. Optical clocks and cavity-stabilized lasers currently set the benchmark for frequency stability and accuracy, outperforming conventional microwave standards by two orders of magnitude both in short and long-term fractional frequency instability\cite{katori_optical_2011,ludlow_optical_2015}. However, translating the extraordinary stability of these optical systems to more accessible RF frequencies, such as 10~MHz and 100~MHz, poses unique challenges.
Optical clocks operate at frequencies in the hundreds of terahertz and achieve fractional frequency stabilities below $10^{-16}$ on short integration times. This remarkable precision will underpin the redefinition of the SI second \cite{dimarcq_roadmap_2024} and extend the application of optical systems beyond their intrinsic domain. The optical frequency comb (OFC) is central to this effort because it enables phase-coherent division of optical frequencies into the RF and microwave regimes with an exceptional level of spectral purity and stability \cite{jones_carrier-envelope_2000, fortier_generation_2011,xie_photonic_2017}. While previous developments have mainly focused on generating 10~GHz signals \cite{nakamura_coherent_2020}, there remains a demand for equally stable signals at lower frequencies, such as 10~MHz and 100~MHz, for applications requiring long-term temporal coherence, and high spectral purity. Currently, 10~ MHz signals are widely used as a standard reference frequency in many electronic devices and test instruments, serving as the stable timing source for precise measurements.  Additionally, distributing signals at 10~MHz and 100~MHz via coaxial cables is more convenient due to their low loss compared to microwave signals. Nonetheless, transferring stability from optical to RF regimes is limited by noise in photodiodes, quantum noise, thermal effects, and subsequent frequency divider noise.

This paper demonstrates a system that transfers cavity-stabilized laser stability to 100~MHz and 10~MHz signals with short-term instability levels below $10^{-15}$ and $10^{-14}$ respectively. This system will be capable of transferring optical clock stability to domains that are critical for future scientific and industrial applications \cite{yao_optical-clock-based_2019}.

\begin{figure}[h]
	\centering
    \includegraphics[width=\columnwidth]{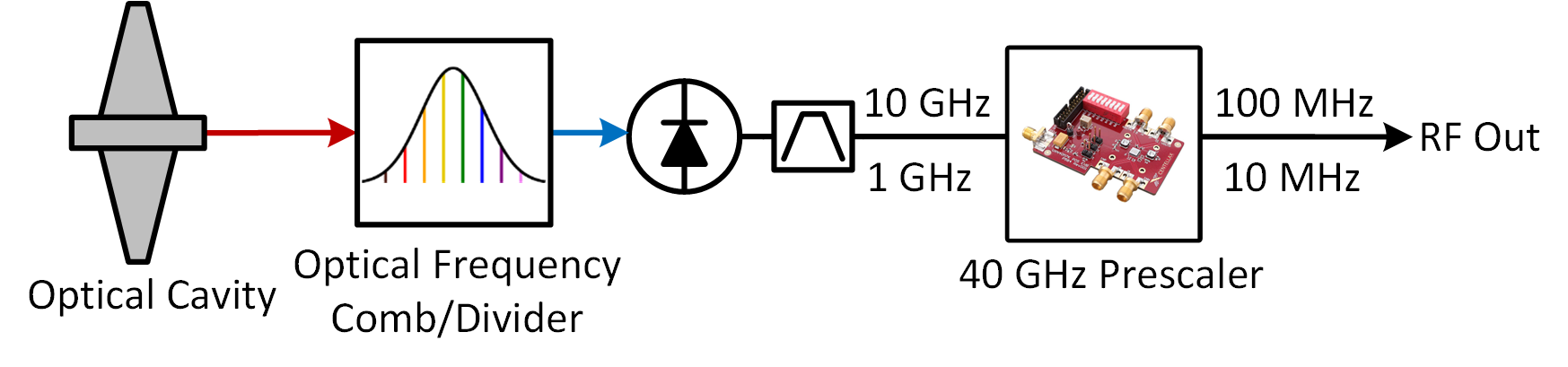}
	\caption{Block diagram of the optical-to-RF synthesis chain.}
	\label{fig:Opt-to-RF Block}
\end{figure}

\section{\label{sec:level1}Description of Optical-to-Radio Frequency Synthesis}
Figure~\ref{fig:Opt-to-RF Block} shows the block diagram of the optical-to-RF synthesis scheme. It employs a high-finesse cavity-stabilized laser, an OFC, and a microwave prescaler. We constructed two separate synthesis chains: one based on a cryogenic silicon cavity-stabilized laser at 1542~nm \cite{matei_15_2017}; the other utilized an ultra-low-expansion (ULE) glass cavity at 1157~nm \cite{schioppo_ultrastable_2017}. Each respective cavity output was converted to a 10~GHz signal via Ti:Sapphire and Er/Yb:glass OFCs \cite{fortier_generation_2011,nardelli_optical_2023}. The 10~GHz microwave outputs were subsequently divided to 100~MHz with commercial prescalers. 

\subsection{\label{sec:level2}Cavity Stabilized Lasers}
The two cavity-stabilized lasers used in the experiment were originally designed and built to probe ultra-narrow atomic resonances in $^{171}$Yb \cite{mcgrew_atomic_2018} and $^{87}$Sr \cite{aeppli_clock_2024} optical lattice clocks. These lasers exhibit coherence times of up to several seconds and ultra-low thermally limited phase noise. To achieve such high performance, the lasers are phase-stabilized to Fabry-Perot cavities via a Pound-Drever-Hall (PDH) lock, whereby the length stability of the cavities is transferred to the frequency and phase of the light.

The 1157 nm laser, used to probe the $^{171}$Yb clock, is based on a ULE glass cavity at room temperature with a finesse of 877,000. The light is frequency-doubled to reach the clock transition at 578 nm. The 1542 nm laser used to probe the $^{87}$Sr clock is based on single-crystal silicon designed to operate at 124 K with a finesse of 500,000. A frequency comb is used to transfer stability to the Sr clock transition frequency at 698 nm. Both optical reference cavities are protected by several layers of thermal isolation, as well as active and passive vibration isolation. The cavities are engineered to neither expand nor contract (zero coefficient of temperature expansion) at their operating temperatures. The cavities are operated at a temperature at which the coefficient of thermal expansion is nominally zero.

\subsection{\label{sec:level2}Optical Frequency Combs}
We generated the two ultrastable 10~GHz signals for our synthesizer using two optical frequency combs to divide down the optical references. One comb was based on a Ti:Sapphire mode-locked laser producing pulses at a rate of 1~GHz \cite{fortier_generation_2011}, and the other comb was based on an Er/Yb:glass mode-locked laser with a 500~MHz repetition frequency ($f_{rep}$) \cite{nardelli_optical_2023}. Both combs were fully stabilized with one comb tooth phase-locked to the optical reference laser, $v_\text{opt}$, while simultaneously stabilizing the carrier offset frequency, $f_{o}$, detected via a \mbox{f-to-2f} interferometer. This transferred the optical reference stability to the comb mode spacing, $f_{rep} = \frac{(v_{\text{opt}}-f_{o})}{n}$, where $n$ is an integer on the order of $2\times10^{5}$. A photodetector converted the laser pulse train to a microwave signal comb with 1~GHz spacing \cite{li_high-power_2011}. A bandpass filter (BPF) then selected the desired harmonic of $f_{rep}$, either 10~GHz or 1~GHz. 

In the absence of added photodetector noise, photonically generated microwave signals permit a reduction in the phase modulation (PM) spectral density noise of the optical reference by $(n/m)^2$ when photodetecting the $m$\textsuperscript{th} harmonic of the OFC repetition rate. When dividing the 1157~nm (259~THz) and 1542~nm (195~THz) optical reference to 10~GHz, this results in a reduction in phase noise by 88~dB and 86~dB, respectively. The phase noise of 10~GHz and 1~GHz signals, scaled to 100~MHz and 10~MHz, is shown later in Section \ref{Section:Results}.

\subsection{\label{sec:level2}Prescalers}
We used commercial digital dividers (Microchip Prescaler: UXN40M7KE), which are specified for input frequencies between 500~MHz and 40~GHz with integer division ratios between 1 and 127 \cite{noauthor_uxn40m7k-prescaler_nodate}. Digital frequency dividers generally support wideband operation with a compact form factor but tend to exhibit relatively high residual phase modulation (PM) noise. In contrast, analog regenerative frequency dividers (RFD) can outperform digital designs in phase noise, but usually offer narrower operating bandwidths and require careful optimization \cite{miller_fractional-frequency_1939,rubiola_phase_1992,gupta_novel_nodate,hati_ultra-low-noise_2012}. The digital divider we employed here proved to have very low residual phase noise at offset frequencies below 1~kHz, and showed an exceptionally high stability at output frequencies of 100~MHz.

\section{\label{sec:level1}Measurement Techniques}
\subsection{\label{sec:level2}Residual Phase Noise and Frequency Stability Measurements of Prescalers}
In previous work\cite{hati_30_2022}, we used these microwave prescalers to generate reference signals for characterizing the instability of 30~GHz divide-by-2 regenerative dividers. However, the result was limited to a fractional frequency instability of $\sigma_y(1~\text{s})=1\times10^{-15}$ due to measurement system noise floor, even after several days of averaging. These results implied that the residual instability of these digital prescalers was likely below $10^{-15}$ at 1~s but remained beyond measurement capability using the available digital measurement systems. We also attempted a conventional analog cross-spectrum PM noise measurement \cite{walls_cross-correlation_1992} but due to high residual amplitude modulation (AM) noise present in the prescalers (see Figs.~\ref{fig:100 MHz PM&AM} and \ref{fig:10 MHz PM&AM}), the measurement was affected by AM-to-PM conversion in the phase detectors.

\begin{figure}
	\centering
    \includegraphics[width=\columnwidth]{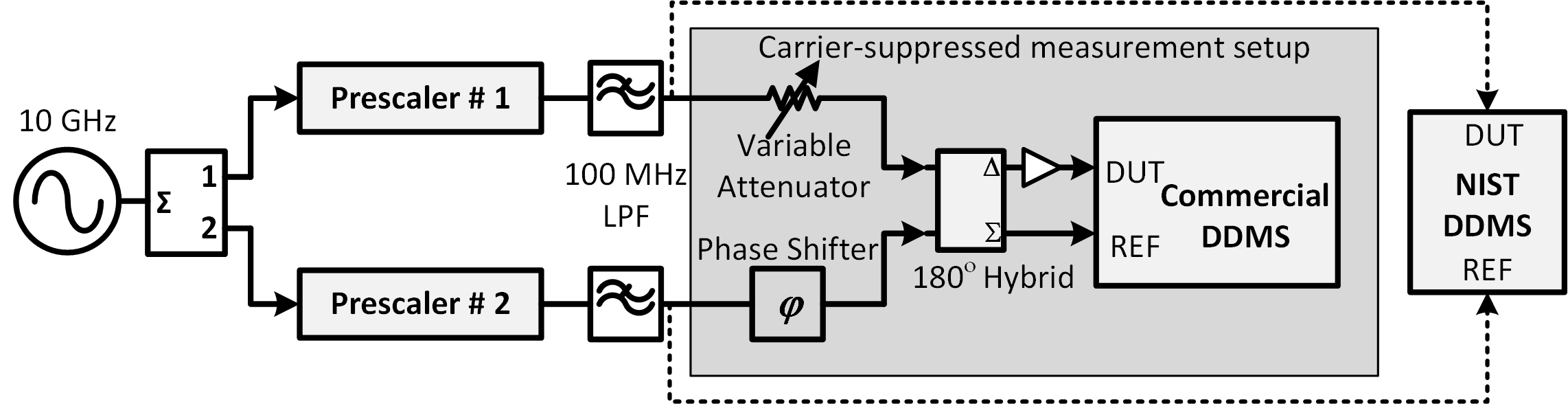}
	\caption{Block diagram of the residual measurement schemes used to evaluate a pair of dividers. DDMS: Direct Digital Measurement System.}
	\label{fig:Residual MS}
\end{figure}

\subsubsection{\label{sec:level3}Carrier-Suppression Technique}
To bypass these limitations, we implemented a carrier-suppression (CS) measurement scheme~\cite{sann_measurement_1968, ivanov_microwave_1998, hati_ultralow_2024} for evaluating a pair of prescalers, as shown in Fig.~\ref{fig:Residual MS}. By summing two phase-aligned signals at the $\Sigma$-port and subtracting them at the $\Delta$-port of a $180^\circ$ hybrid, we increased the effective sensitivity in phase noise measurements proportionally to the amount of carrier suppression achieved. This method is very effective for residual measurements, but it can be challenging for making absolute measurements where a phase locked loop (PLL) is necessary to maintain the phase relationship required for  carrier suppression. The frequency response of the PLL suppresses the measurement of longer-term frequency fluctuations, and thus the technique is not ideal for extended averaging of non-residual measurements.

\subsubsection{\label{sec:level3}Direct Digital Measurement System}
We used a novel and enhanced performance multichannel direct digital measurement system (DDMS) currently under development at NIST \cite{pomponio_oscillator_2025,pomponio_flicker_2025}. The DDMS  can support up to eight inputs (or channels), however, for this measurement campaign only four inputs were used. By measuring and correcting for aperture jitter, voltage reference and residual flicker noise of the analog-to-digital converters (ADCs), this new DDMS achieves more than a 35~dB reduction in close-to-the-carrier residual phase and amplitude noise, and more than a 20-fold reduction in residual Allan deviation ($\sigma_y(\text{t})$) compared to our previously developed system \cite{pomponio_direct_2024} at 100~MHz carrier. Under ideal conditions, the new DDMS exhibits a single channel residual noise of $-147$~dBc/Hz at 1~Hz offset, with a flicker-corner of about 30~mHz and a residual frequency stability of $1.3\times10^{-16}$ at a 1~s averaging time for 100~MHz carriers. Both frequency and time domain performance can be further improved through cross-correlation averaging. For example, with a full-scale input signal of +9~dBm at each input, the DDMS 100~MHz noise floor improves to near or below $1\times10^{-17}$ at 1~s for a  0.5~Hz measurement bandwidth \cite{pomponio_flicker_2025}. Additionally, the AM-to-PM isolation has been verified to exceed 40~dB at a 1~Hz offset. This capability has enabled unprecedented phase noise and instability measurements of these prescalers at 10~MHz and 100~MHz outputs in both residual and absolute configurations.

\begin{figure}
	\centering
    \includegraphics[width=\columnwidth]{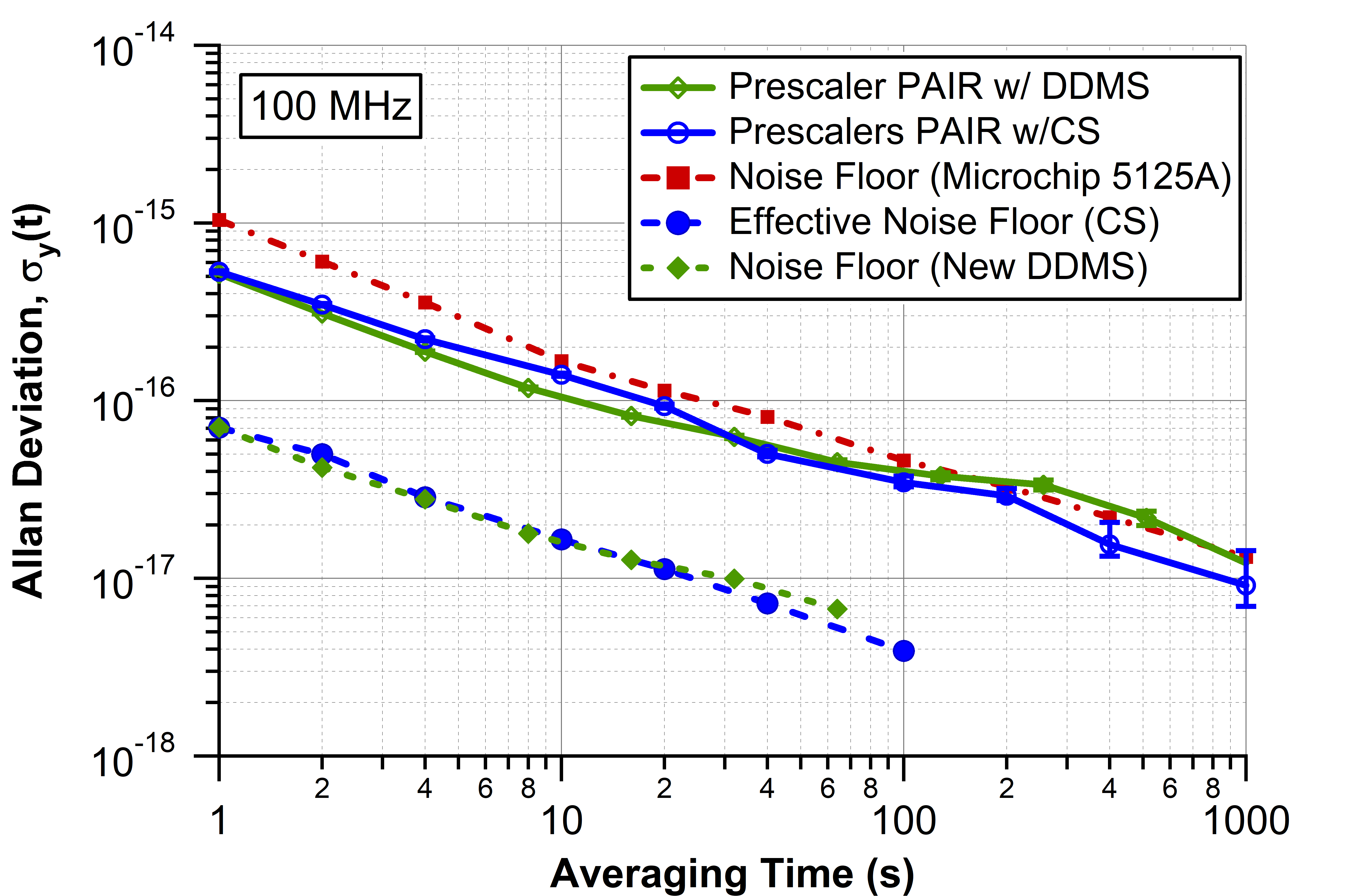}
	\caption{Residual fractional frequency instability of a pair of prescalers at 100~MHz measured with the carrier-suppression (CS) and direct digital techniques. Noise floor of carrier-suppression and both commercial and NIST-developed digital measurement systems are also shown. Confidence interval of error bars = 1~sigma, and measurement bandwidth = 0.5 Hz.}
    \label{fig:Noise Floor}
\end{figure}

We first performed a residual frequency instability comparison of a pair of prescalers, dividing 10~GHz down to 100~MHz, using both the carrier-suppression method and our new digital system. As shown in Fig.~\ref{fig:Noise Floor}, the results agreed closely. Unlike the commercial system,  the measurement floors of both the carrier-suppression system and the new digital system were sufficiently low to reveal the prescalers' inherent noise. Because the prescalers only provide output power of about +1~dBm, the cross-correlated frequency stability floor was slightly degraded, but remained well below $10^{-16}$ at 1~s. The noise floor of the commercial, CS, and digital measurement systems was measured using a common signal across all channels and at roughly identical power levels as those used for the Allan deviation, absolute, and residual noise measurements.

We also discovered that individual prescalers exhibited up to 6~dB variation in their PM noise, AM noise, and frequency stability performance. To isolate the residual performance of the best prescaler, we used a three-divider approach, also known as the cross-spectrum (or cross-correlation) three-corner hat method \cite{fest_individual_1983}. In this configuration, a common signal `y' generated from the Er/Yb:glass OFC (Fig.~\ref{fig:Res-Ab Block}a) was used to drive all three prescalers - the device under test (DUT), and two reference prescalers (REF-1 and REF-2)  as shown in Fig.~\ref{fig:Res-Ab Block}b. This approach enhances sensitivity of noise measurements by reducing uncorrelated noise from the two reference prescalers by $\sqrt{k}$, where $k$ is the number of FFT averages. This allows an accurate determination of residual phase noise and frequency instability of a single prescaler, the DUT.

\subsection{\label{sec:level2}Absolute Phase Noise and Frequency Stability Measurements of Optical-to-RF Synthesizers}
\begin{figure}
	\centering
    \includegraphics[width=\columnwidth]{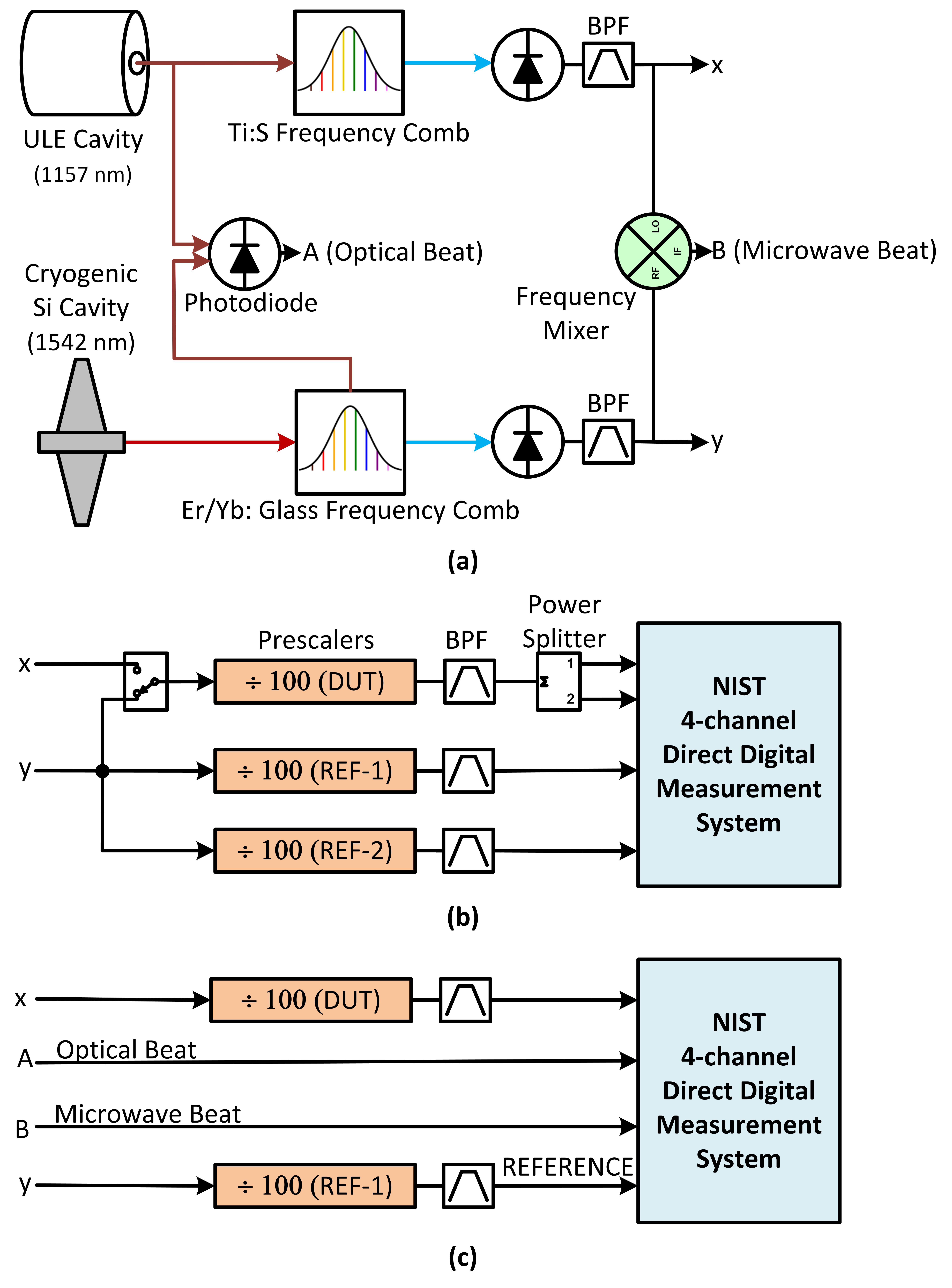}
	\caption{Block diagram illustrating measurement of residual noise, absolute noise, and fractional frequency instability of the optical-to-RF synthesis chain. (a) Optical-to-microwave synthesis, and optical beat and microwave beat generation. (b) Microwave-to-RF generation and set-up for residual and absolute noise measurements. For residual measurements, all prescalers were connected to a common signal `y' and for absolute measurements, the prescaler `DUT' was connected to `x' and two reference prescalers were connected to `y'. A physical switch was not used, it's shown in this block diagram to illustrate the conversion between residual to absolute measurements. (c) Set-up for fractional frequency fluctuations measurement. Signals `A', `B', and `x$\div100$',  were all measured with respect to REFERENCE signal.  }
	\label{fig:Res-Ab Block}
\end{figure}

We next characterized two optical-to-RF synthesizer chains (Fig.~\ref{fig:Res-Ab Block}a). The cross-spectrum scheme in Fig.~\ref{fig:Res-Ab Block}b measured the absolute phase noise and frequency instability between the two independently generated optical signals, each divided down to 100~MHz. The reference prescalers (REF-1 and REF-2) were connected to a common microwave signal (`y'), while the DUT was driven by the other microwave signal (`x'). This approach measured the combined noise from both cavity-stabilized lasers, both OFCs, their photodiodes, and only one prescaler (the DUT).  

Measurements were performed at three points along the synthesis chain. Optical stability was assessed by measuring the heterodyne beat between two cavity-stabilized lasers generated via the Er/Yb:glass frequency comb (output A). Microwave frequency stability was assessed by measuring the heterodyne beat between two 10 GHz signals (output B). Finally, the stability of the 100 MHz prescaler outputs (derived from 10 GHz) was measured directly using a digital system, without heterodyne mixing. A single photodetector and a frequency mixer generated the heterodyne difference frequency for the optical and microwave signals respectively, and the resulting beat signals were also analyzed with the new DDMS.   
We additionally conducted similar measurements at 10~MHz by dividing the 1~GHz frequency comb signal by 100. 

\section{Results} \label{Section:Results}
\begin{figure}
	\centering
    \includegraphics[width=\columnwidth]{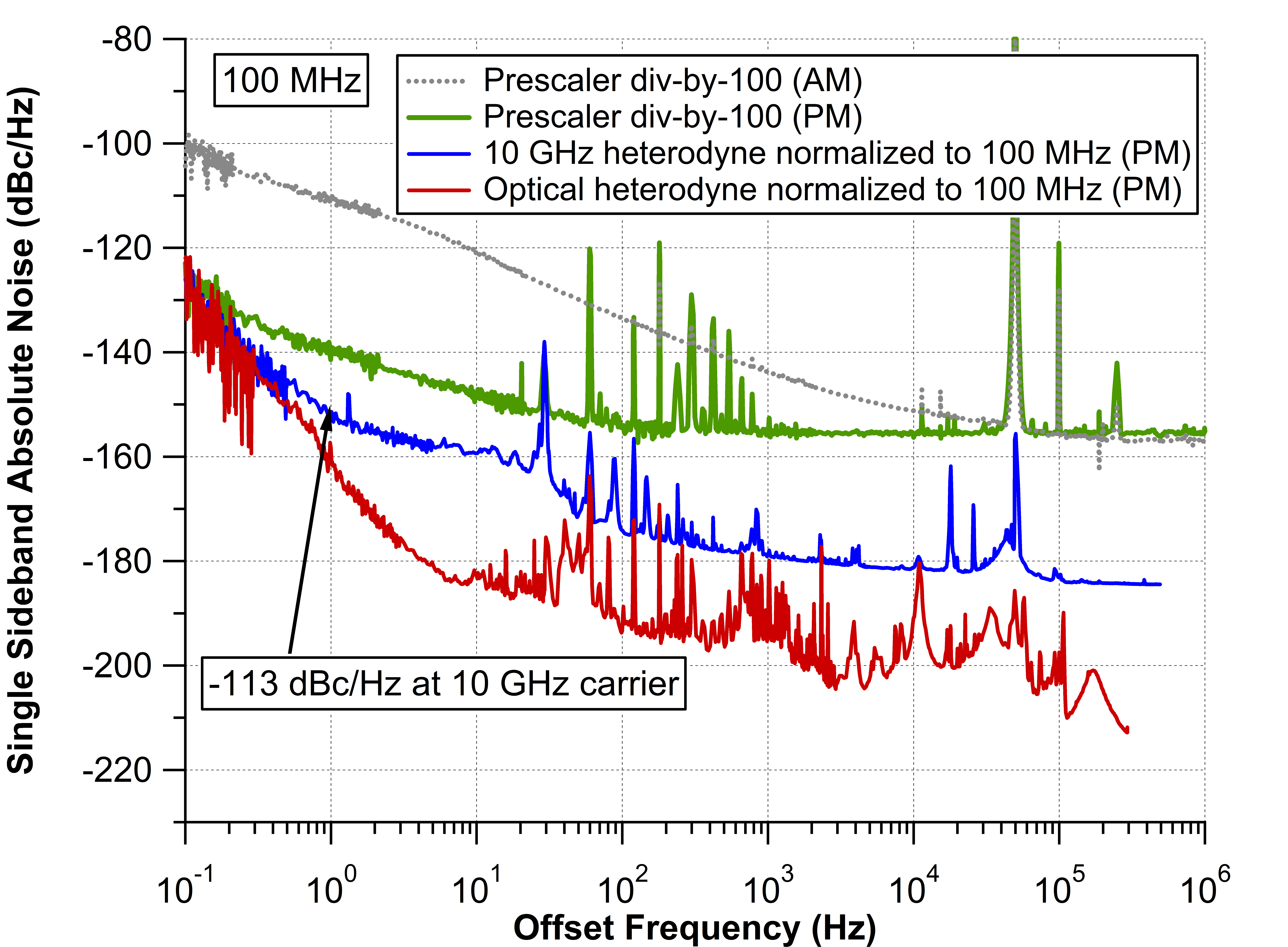}
	\caption{Absolute phase noise of optical, microwave (10~GHz) and 100~MHz prescaler output signals. All phase noise plots are normalized to 100~MHz. The AM noise (in gray) is dominated by the prescaler AM noise.}
	\label{fig:100 MHz PM&AM}
\end{figure}

We evaluated the prescalers at two input frequencies and a divide ratio of 100: 10~GHz down to 100~MHz, and 1~GHz down to 10~MHz. Fig.~\ref{fig:100 MHz PM&AM} illustrates the absolute phase noise and AM noise at 100~MHz, as well as the phase noise of the  optical heterodyne and 10~GHz microwave heterodyne, which are all normalized to 100~MHz. At a 1~Hz offset, the scaled 10~GHz beat had a phase noise of about -153~dBc/Hz which corresponds to -113~dBc/Hz at 10~GHz, while the 100~MHz signals achieved approximately -140~dBc/Hz. For offsets above 0.2~Hz, the phase noise of 100~MHz signal was dominated by the prescaler noise. The prescaler’s AM noise was roughly 30~dB higher than its PM noise at 1~Hz offset, which could be problematic with standard analog cross-spectrum techniques due to AM-to-PM conversions at the phase detectors. 

\begin{figure}
	\centering
    \includegraphics[width=\columnwidth]{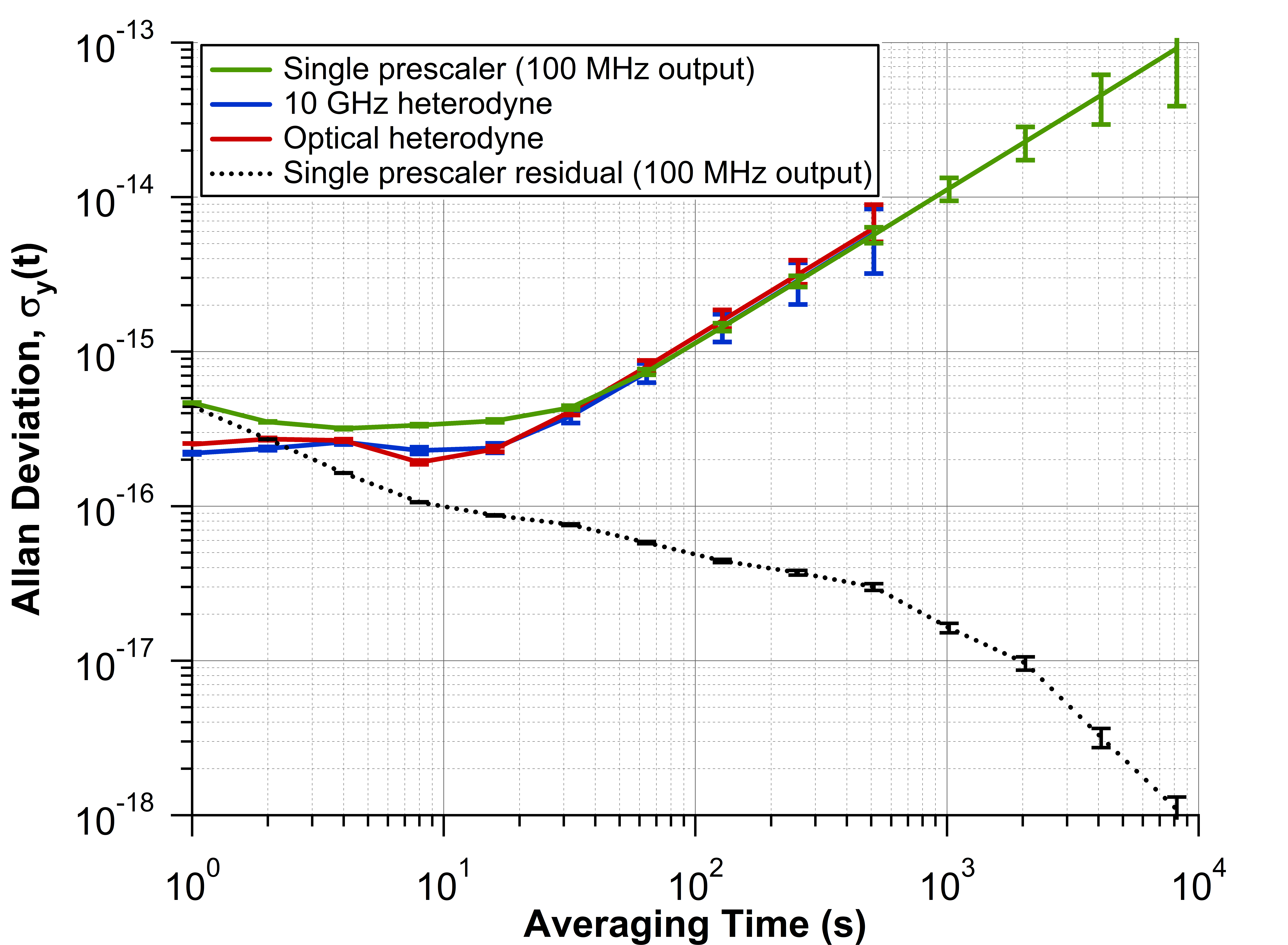}
	\caption{Fractional frequency instability of optical, microwave (10~GHz), and 100~MHz signals. It shows the prescaler can transfer the stability of optical signal nearly perfectly without degradation above 20~s. Please note that the single prescaler result is obtained via cross-covariance method. The prescaler demonstrates residual instability of $4.7\times10^{-16}$ at 1~s and  approaches $10^{-18}$ at longer averaging time. Confidence interval of error bars = 1~sigma, and measurement bandwidth = 0.5~Hz.}
	\label{fig:100 MHz ADEV}
\end{figure}

\begin{figure}
	\centering
    \includegraphics[width=\columnwidth]{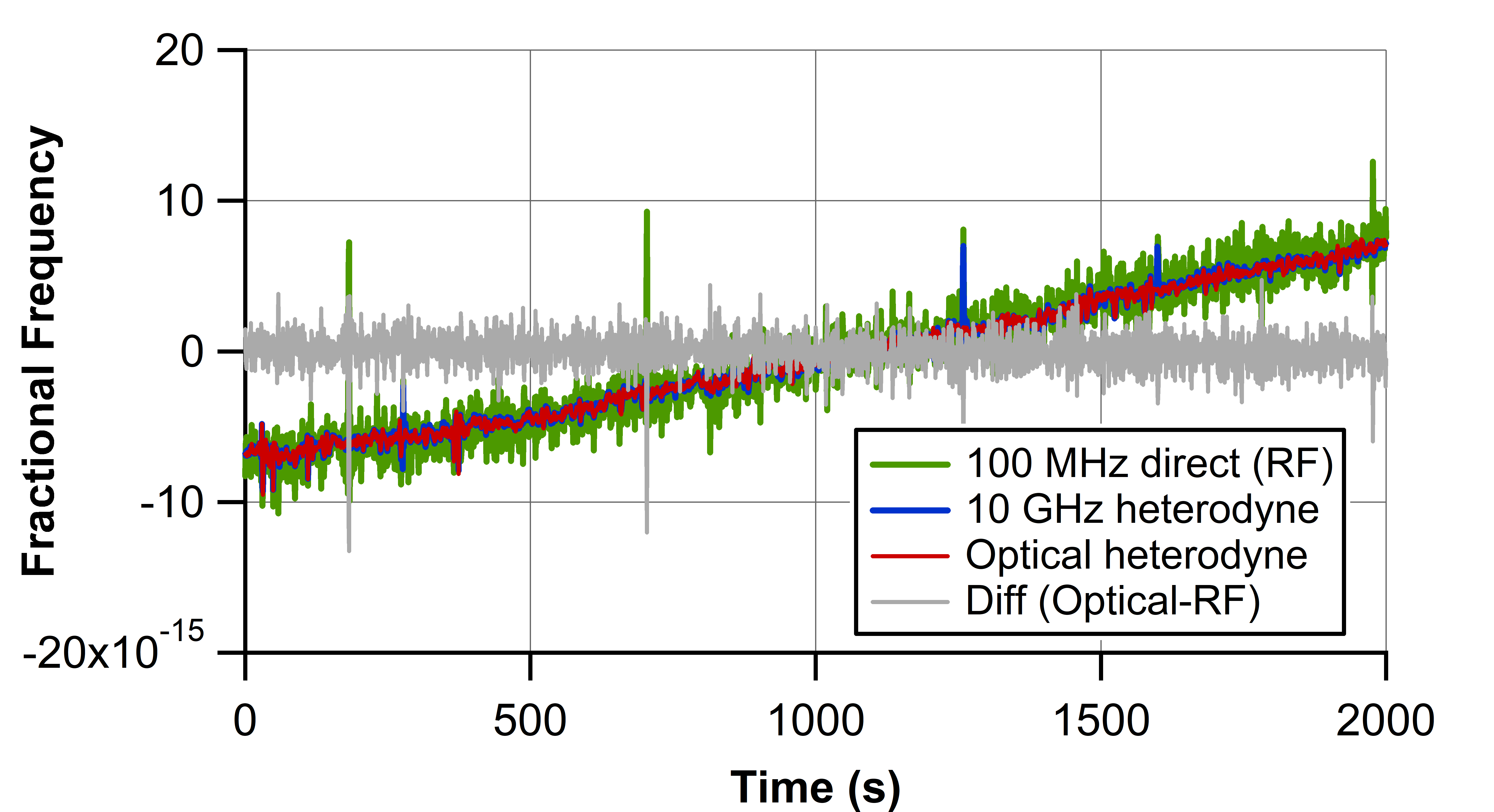}
	\caption{Fractional frequency fluctuations between two 100~MHz signals from a pair of prescalers, 10~GHz microwave beat, and   optical beat.}
	\label{fig:100 MHz FF}
\end{figure}

We also calculated Allan deviation statistics to quantify the frequency instability of the 100 MHz and both heterodyne beats as shown in Fig.~\ref{fig:100 MHz ADEV}. The synthesis achieves $4.7\times10^{-16}$ absolute instability at 1~s for the 100~MHz signal. This represents the first demonstration of such low instability at 100~MHz, and significantly surpassing the performance of earlier system\cite{hati_state---art_2013}. The prescaler alone supported $10^{-18}$ residual stability above $8{,}000$~s when we used the cross-covariance method \cite{fest_individual_1983} to remove the reference prescalers' noise.
Fig.~\ref{fig:100 MHz FF} shows the real-time fractional frequency fluctuations of the optical beat, the microwave beat, and two 100~MHz outputs of a pair prescalers over a $2{,}000$~s interval. These signals clearly show the differential drift between the two cavities. Even if the difference between the optical and the 100~MHz signals showed fluctuations, due to the prescaler residual noise, it remained within the $\pm 4\times10^{-15}$ range if the few glitches are ignored. The observed frequency drift was primarily attributed to the ULE cavity, which has a feed-forward drift compensation that was not optimized during these measurements \cite{schioppo_ultrastable_2017, milner_demonstration_2019}. The configuration depicted in Fig.~\ref{fig:Res-Ab Block}c was used for the evaluation of fluctuations in fractional frequency. 

\begin{figure}
	\centering
    \includegraphics[width=\columnwidth]{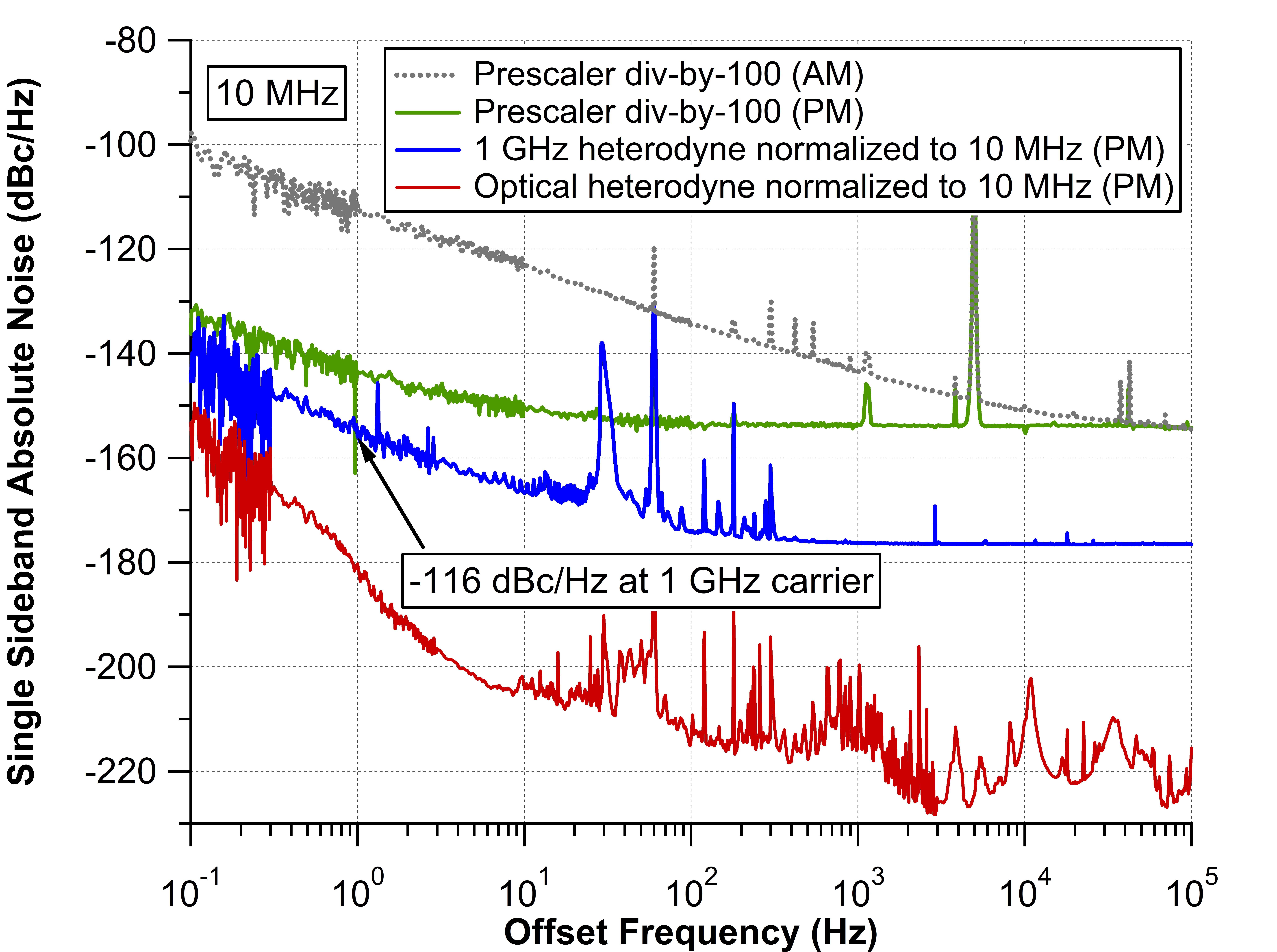}
	\caption{ Absolute phase noise of optical, microwave (1~GHz) and 10~MHz prescaler output signals. All phase noise plots are normalized to 10~MHz. The AM noise (in gray) is dominated by the prescaler AM noise.}
	\label{fig:10 MHz PM&AM}
\end{figure}

\begin{figure}
	\centering
    \includegraphics[width=\columnwidth]{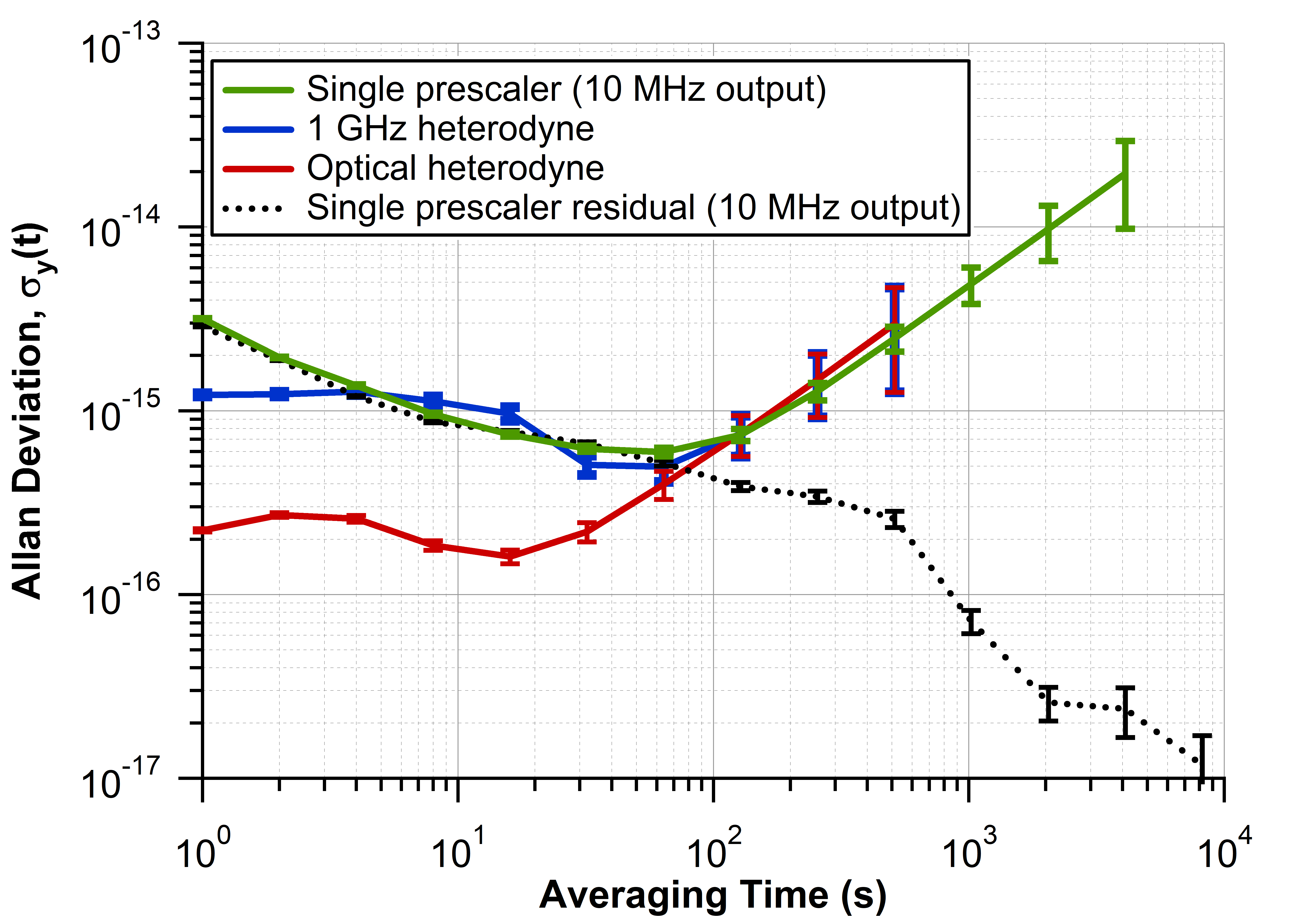}
	\caption{Fractional frequency instability of optical, microwave (1~GHz), and 10~MHz signals. It shows the prescaler can transfer the stability of 1~GHz OFC signal nearly perfectly without degradation above 3~s. Please note that the single prescaler result is obtained via the cross-covariance method. The prescaler demonstrates residual instability of $\approx3\times10^{-15}$ at 1~s and  approaches $10^{-17}$ at longer averaging times. Confidence interval of error bars = 1~sigma, and 
 measurement bandwidth = 0.5~Hz.}
	\label{fig:10 MHz ADEV}
\end{figure}

\begin{figure}
	\centering
    \includegraphics[width=\columnwidth]{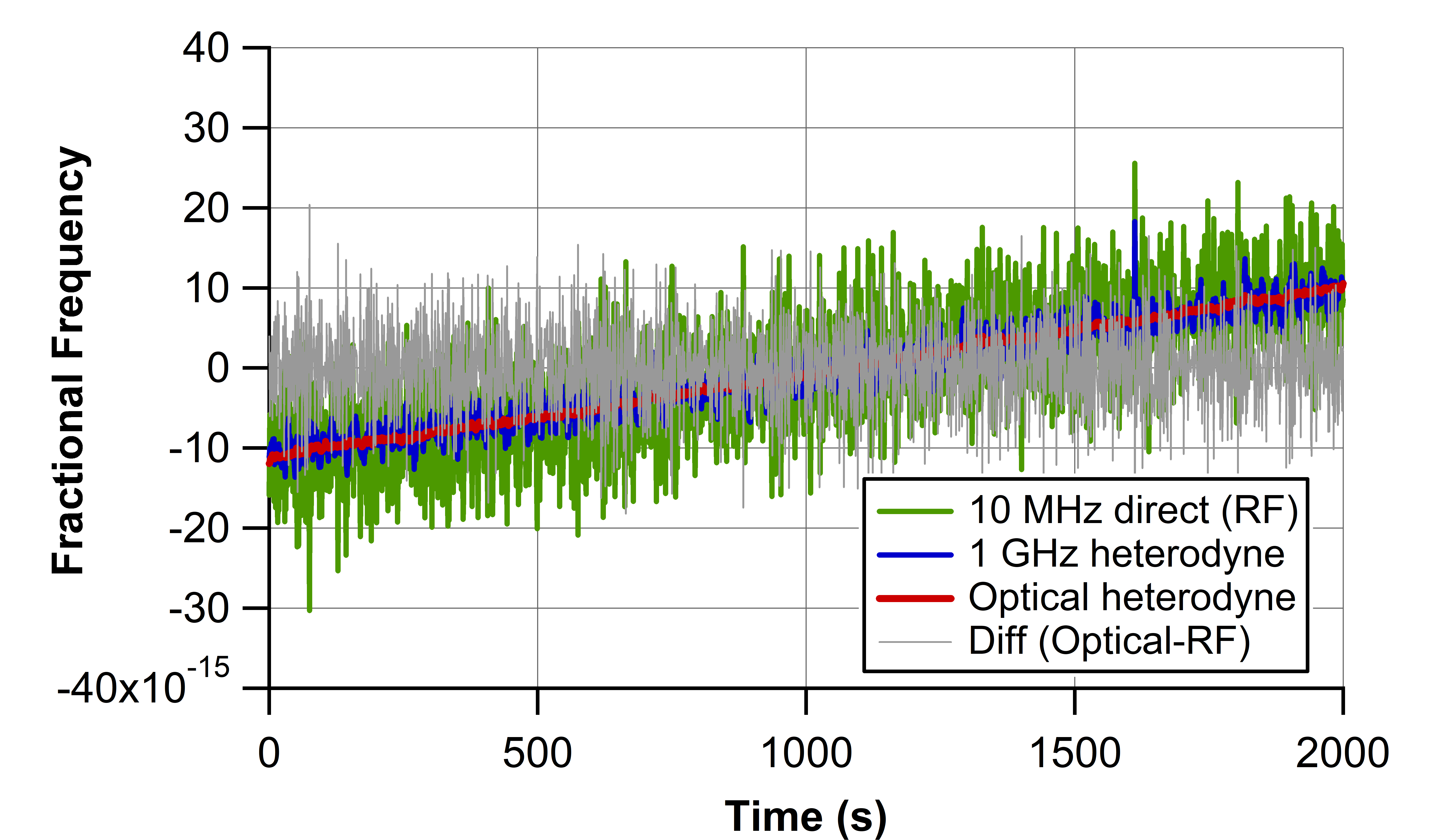}
	\caption{Fractional frequency fluctuations between two 10~MHz signals from a pair of prescalers, 1~GHz microwave beat, and optical beat over 2,000~s.}
	\label{fig:10 MHz FF}
\end{figure}

We repeated the above measurements with a microwave heterodyne between two 1~GHz signals (output B), and a direct digital measurement of the 10~MHz prescalers' output (divided down from 1~GHz). The phase noise of the 1 GHz signal does not follow the theoretical noise scaling by $n^2$ from the optical. As shown in  Fig.~\ref{fig:10 MHz PM&AM}, at a 1~Hz offset, the 1~GHz signal has a phase noise of about -116~dBc/Hz which is only 3 dB lower than the phase noise of 10 GHz signal. This may be due to the photodiode's high flicker noise and/or high relative intensity noise (RIN) of the laser, which can result in excessive phase noise due to AM-to-PM conversion. For the 1~GHz to 10~MHz division, the prescaler’s flicker noise contribution was comparatively larger at lower offset frequencies than at 100 MHz output, about -144~dBc/Hz at 1 Hz. 
Furthermore, the 10~MHz signal exhibited an absolute fractional  instability of $\approx3\times10^{-15}$ at 1~s, for a single prescaler (Fig. \ref{fig:10 MHz ADEV}). Beyond 100~s of averaging, however, the prescaler’s impact was negligible, and the absolute instability of the 10 MHz signal approached that of the optical reference. The prescaler demonstrates residual instability of $\approx3\times10^{-15}$ at 1~s and  approaches $10^{-17}$ at longer averaging times. The difference of fractional frequency between the optical and the 10~MHz signals showed fluctuations within $\pm 20\times10^{-15}$ as shown in Fig. \ref{fig:10 MHz FF}.

\begin{figure}
	\centering
    \includegraphics[width=\columnwidth]{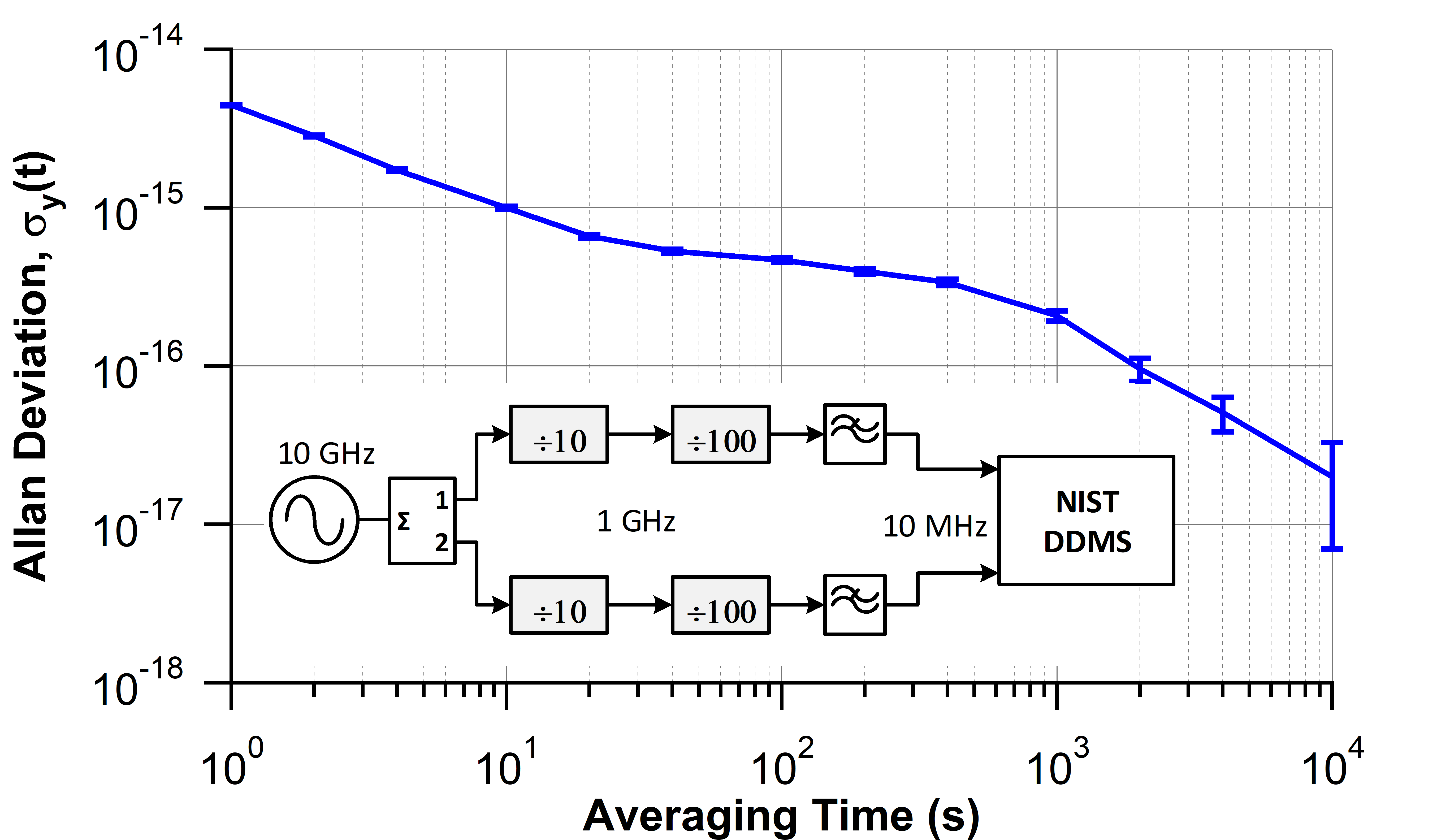}
	\caption{Fractional frequency instability of a pair of cascaded prescalers dividing 10~GHz to 10~MHz. Confidence interval of error bars = 1 sigma, and measurement bandwidth = 0.5 Hz.}
	\label{fig:ADEV div by 1000}
\end{figure}

In addition, multiple prescalers can be used in series for higher division factors. For example, a 1,000 division from 10~GHz to 10~MHz demonstrates a fractional instability of better than $5\times10^{-15}$ at 1~s for two cascaded stages, shown in Fig.~\ref{fig:ADEV div by 1000}. In this configuration, the output noise of the first prescaler in the cascade is reduced by $n_2^2$, where $n_2$ is the frequency division ratio of the second stage. Therefore, for higher values of $n_2$, the output noise contribution is dominated by the second stage prescalers.

Frequency instability measurements at this level is highly sensitive to environmental effects.  Long-term residual and absolute measurements were performed at night or on weekends to prevent vibration induced disturbances. Although, the microwave-to-RF synthesis sections  were sensitive to vibration and temperature, no stabilization was used other than shielding the prescalers from direct airflow from the heating, ventilation, and air conditioning (HVAC) system.

\begin{figure}
	\centering
    \includegraphics[width=\columnwidth]{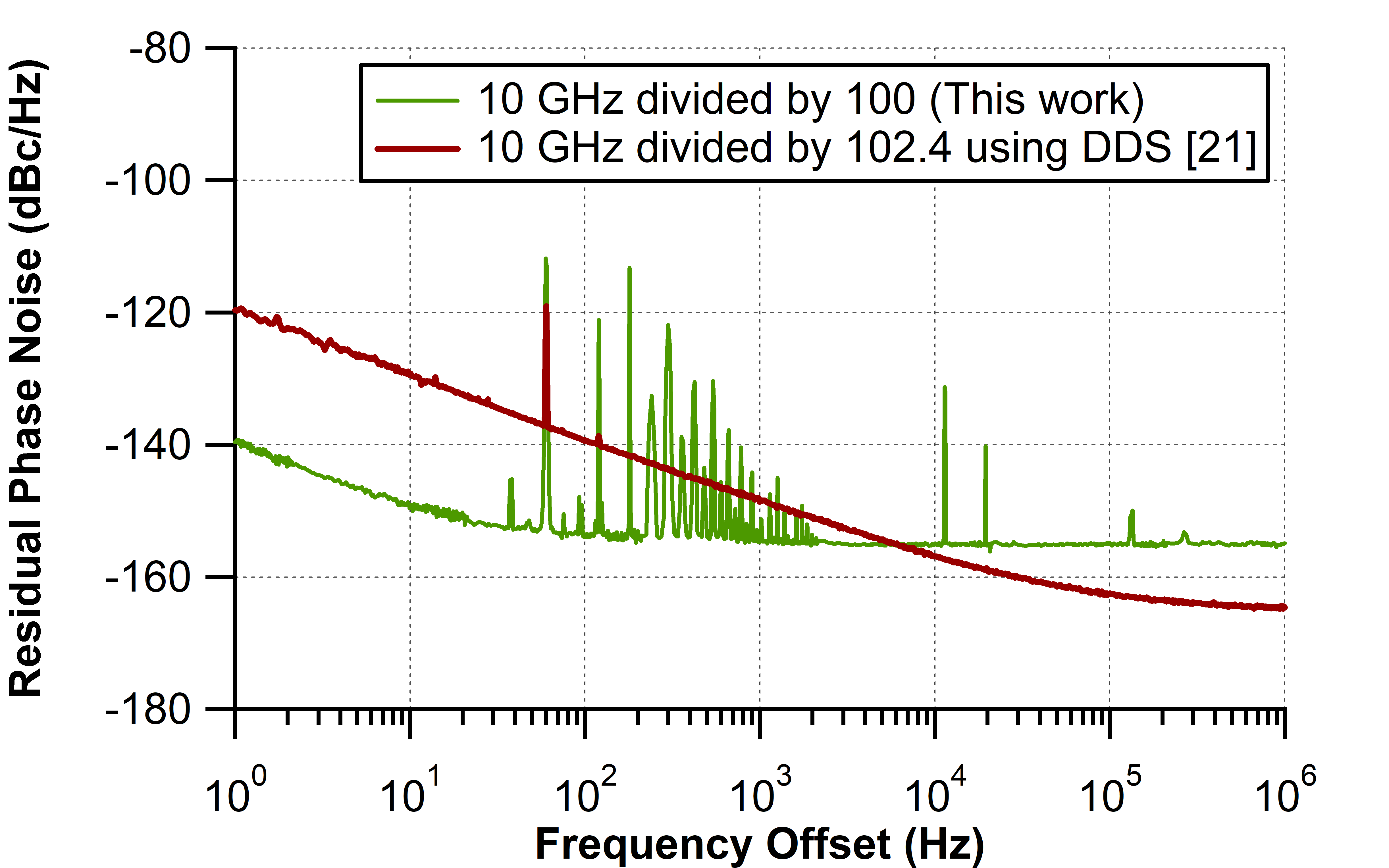}
	\caption{Comparison of residual phase noise of digital dividers for a frequency division factor of about 100 for a carrier frequency of 10~GHz.}
	\label{fig:AD dividers}
\end{figure}

Finally, Fig.~\ref{fig:AD dividers} compares the residual noise of this commercial prescaler with that of a typical DDS \cite{pomponio_ultra-low_2024} for a 10 GHz input frequency. The prescaler we employed exhibits phase noise that is almost 20~dB lower than the DDS at a 1~Hz offset for a frequency division factor of about 100. 

\section{Conclusion}
We presented frequency division from the optical domain down to 100~MHz with an absolute instability of $4.7\times10^{-16}$ at 1~s, corresponding to a phase noise of -140~dBc/Hz at a 1~Hz offset. To our knowledge, this is the first demonstration of 100~MHz signals achieving this level of stability, surpassing previous approaches of optical-to-RF synthesis using regenerative frequency dividers \cite{hati_ultra-low-noise_2012}, parallel DDS technique \cite{pomponio_ultra-low_2024}, microwave frequency synthesizer \cite{yan_photonic_2018} or generation directly from cryogenic oscillators \cite{calosso_frequency_2019, al-ashwal_low_2017}. We achieved this performance utilizing ultra-high stability commercial frequency prescalers. This is also the first time that 100~MHz measurements, at this level, were performed directly without utilizing a heterodyne beat to increase sensitivity using a newly developed multi-channel digital measurement system with state-of-the-art level performance. This system exhibits a single channel residual noise of $-147$~dBc/Hz at 1~Hz offset, and a residual frequency stability of $1.3\times10^{-16}$ at a 1~s averaging time for 100~MHz carriers.
Additionally, 10~MHz RF signals were generated from the optical domain, and we observed absolute fractional instabilities on the order of $\approx3\times10^{-15}$ at 1~s, dominated by prescaler noise.  
This study finds that these high performance prescalers can transfer the pristine stability of the optical clocks to usable RF frequencies and will be able to facilitate ultrastable frequency references for future precision metrology and timing systems.

\begin{acknowledgments}
The authors thank Vladislav Gerginov, and Nazanin Hoghooghi for their constructive feedback on this manuscript.
Certain equipment, instruments, software, or materials are identified in this paper to specify the experimental procedure adequately. Such identification is not intended to imply recommendation or endorsement of any product or service by NIST, nor is it intended to imply that the materials or equipment identified are necessarily the best available for the purpose.
\end{acknowledgments}
\section*{Data Availability Statement}
The data that support the findings of this study are available
from the corresponding author upon reasonable request.
\section*{References}
\bibliography{references_local}

\end{document}